\documentstyle[fleqn]{article}
\title{5-Dimensional Covariance and Generation
of Solutions of Einstein Equations}
\author{Sergey S.Kokarev\thanks{sergey@yspu.yar.ru}}
\date{Department of theoretical physics, r.409, YSPU,
                       Respublikanskaya 108, Yaroslavl, 150000, Russia}

\begin{document}

\maketitle

\begin{abstract}
\small A generation procedure, based on the 5-dimensional covariance
of the Kaluza-Klein theory, is developed.
The procedure allows one to obtain exact
solutions of the 4-dimensional Einstein equations
with electromagnetic and scalar fields
from vacuum 5-dimensional solutions
using special 5-dimensional coordinate transformations.
Relations between the physical properties of the
resulting solutions and invariant geometrical properties of
the generating Killing vectors are found out.
\end{abstract}

\section{Introduction}
\label{gen4}

Searches for exact solutions of the Einstein equations
are stimulated by the internal logic of General Relativity (GR).
A special role in this problem is played by the so-called generation
methods which give new solutions without solving the Einstein equations. As
a rule, they are based on some original solution and special kinds of
transformations of the metric, dynamical Lagrangian variables and coordinates
\cite{exs}.

In the present paper a generation method based on the classical version of
the Kaluza-Klein theory (KKT) is worked out.  Theories of the Kaluza-Klein
type have recently gained a new physical content.  It concerns both
mathematical and physical aspects of the theory, which were out of
consideration earlier \cite{wes1,wes3,kok1}.  The method to be proposed is
related to the well-known property of the KKT, which takes place in any
version of the theory due to its multidimensional covariance:  the
4-dimensional interpretation of multidimensional geometrical objects depends
on the choice of the multidimensional coordinate system
\cite{wes1,vlad1,kok}.  Thus by general coordinate trans\-for\-ma\-ti\-ons
one can generate new properties of 4-dimensional space-times, though the
multidimensional world remains the same.  This fact has been used by  a
number of authors.  For instance, in \cite{ros} the electromagnetic field
generation procedure in 4-D vacuum by a special 5-D coordinate
transformation has been pro\-po\-sed.  In  \cite{vlad1} a linear
transformation in the $(x^{0}x^{5})$-plane generating an exact solutions
with electromagnetic and scalar fields from the Schwarzschild metric has
been used.  In \cite{kok} these transformations have been applied to
Kramer's metric. In \cite{wes1} an explicit form of 5-D coordinate
transformations which transform the flat 5-D metric into a cosmological
metric of Friedmann type is given.  The present approach includes the
methods proposed in \cite{vlad1,kok,ros}, and partially in \cite{wes1}, as
special cases.

The idea of our method has a clear geometric meaning.  We assume (which is
usual in KKT) that the invisibility of the 5th dimension is due to an
isometry of the 5-dimensional space-time.  Then the set of integral curves
of the Killing vector field can be associated with the coordinate lines of
$x^{5}$.  If 5-D space-time admits more than one Killing vector, the
following uncertainty appears: which Killing vector should be connected with
a 5-D coordinate system?  If we first connect it with one Killing vector
field, then, by a suitable coordinate transformation with another one we get
a transition between different 4-D physical worlds.

Section 2 presents a sketch of the classical KKT.

In Section 3 the class of admissible
5-D coordinate transformations is generalized.
In addition, we prove a number of propositions
connecting the invariant properties of Killing vectors with the physical
properties of the resulting 4-dimensional solutions.

In Section 4 the  method is applied to a flat metric.

\section{Kaluza-Klein theory and monad formalism}\label{kal}

The Kaluza-Klein theory is a direct 5-dimensional ge\-ne\-ra\-li\-za\-ti\-on
of GR. It was developed for geometrical unification of gravity and
electromagnetism \cite{kaluza,klein}.  A (1+4)-splitting procedure (monad
formalism) worked out later gives an invariant description
of the fifth dimension and allows one to interpret 5-dimensional objects in
terms of 4-dimensional ones \cite{einb}. A modern presentation of the
classical 5-dimensional KKT to\-ge\-ther with the (1+4)-splitting formalism
can be found in \cite{vlad1}. Some necessary features of the theory are
presented below.

The 5-dimensional metric $G$ defined on a Riemannian space $V_5$ is
the basic geometric object in KKT. It can be taken in the form
\begin{equation}\label{monada}
      G=-\lambda\otimes\lambda+\tilde g
\end{equation}
where $\lambda=\lambda_{A}dx^A$ is the 1-form
determining a (1+4)-splitting of the 5-dimensional space
$(A=0,1,2,3,5)$, $\tilde g$ is the
local metric of a 4-dimensional hypersurface orthogonal to
$\lambda$. These objects satisfy the following conditions:
\[
      \lambda_{A}\lambda^{A}=-1;\ \     \lambda^{A}\tilde g_{AB}=0.
\]
The tilde over $g$ distinguishes the geometric 4-di\-men\-si\-onal metric
and the physical (observable) one (see below).

To shorten the presentation, we make some assumptions
which simplify the mathematical formulae and
``tune" the KKT to the problem of finding exact 4-dimensional solutions%
      \footnote{A general approach is developed in \cite{vlad1}.}.
\begin{enumerate}
\item
The coordinate lines of $x^{5}$
can be associated with a vector field $\vec\lambda$
which corresponds to its special gauge:
\begin{equation}\label{gauge}
\lambda^{A}=\frac{G^{A}_{5}}{\sqrt{-G_{55}}} \ \ \rightarrow \ \
\lambda_{A}=\frac {G_{A5}} {\sqrt{-G_{55}}}.
\end{equation}
\item
We accept the standard KKT argumentation:
the un\-ob\-ser\-va\-bi\-li\-ty of the 5th dimension can be explained
by cylindricity of 5-dimensional world in the fifth coordinate%
      \footnote{The compactification assumption can be accepted as well,
              but here it is not necessary.}.
In the formalism to be presented this means that
$\vec\lambda$  is a Killing vector in $V_5$.
In the gauge chosen the cylindricity condition takes the form
\begin{equation}\label{cyl}
      \partial_5G_{AB}=0.
\end{equation}
\item
The class of admissible coordinate trans\-for\-ma\-ti\-ons
which do not violate (\ref{cyl}) includes
purely 4-di\-men\-si\-o\-nal general coordinate transformations
and gauge transformations:
\[
x^{'5}=x^{5}+f(x^{0},x^{1},x^{2},x^{3}).
\]
\item
Only the 4-dimensional space-time section tensors
projected onto $\vec\lambda$ or orthogonal to $\vec\lambda$
have a physical meaning.
\item
The fifth direction should be spacelike to get a correct
sign of the geometrized energy-mo\-men\-tum tensor of the electromagnetic
field in the right-hand side of the projected 5-dimensional Einstein
equations.  The re\-la\-ti\-on between the 4-vector potential $\vec A$ and
the 4-dimensional part of $\vec\lambda$ has the form
\begin{equation}\label{ident}
      \lambda_{\mu}=(2\sqrt{k}/c^{2})\varphi A_{\mu},
\end{equation}
where $\varphi=\sqrt{-G_{55}}$ is the geometrized scalar field,
$k$ is the Newtonian gravitational constant,
$c$ is the velocity of light and $\mu=0,1,2,3$.
\item
The 5-dimensional Einstein equations are vacuum:
\begin{equation}\label{vacuum}
      {}^{(5)}G_{AB}=0,
\end{equation}
where ${}^{(5)}G_{AB}$ is the 5-dimensional Einstein tensor.
This assumption makes the theory more economic and
close to its modern versions \cite{wes1,kok1}.
\item
The observable 4-dimensional metric is not the projector
$\tilde g$ but the conformally transformed tensor $g$, related to $\tilde g$
by the expression
\begin{equation}\label{conf}
      \tilde g=\varphi^{2}g.
\end{equation}
\end{enumerate}

Taking into account all these items, the projected equations
(\ref{vacuum}) take the form
\begin{equation}\label{e4}
     {}^{(4)}G_{\mu\nu}=\ae T^{\rm (em)}_{\mu\nu}
       +3(\phi_{;\mu;\nu}-\phi_{,\mu}\phi_{,\nu}
      -g_{\mu\nu}(\nabla^{2}\phi + (\nabla\phi)^{2}));
\end{equation}
\begin{equation}           \label{maxw}
      F^{\mu\nu}_{\ \ ;\nu}-3\phi_{,\alpha}F^{\alpha\mu}=0;
\end{equation}
\begin{equation}\label{s4}
      \nabla^{2}\phi+(\nabla\phi)^{2}
            - \frac{1}{6}{}^{(4)}R-\frac{3\ae}{8\pi}F^{2}=0.
\end{equation}
Here:
\medskip
${}^{(4)}G_{\mu\nu}$ is the 4-dimensional Einstein tensor
built up from the observable metric $g$;

\medskip
$F_{\mu\nu}=A_{\nu,\mu}-A_{\mu,\nu}$ is the geometrized electromagnetic field tensor;
\[
      T^{\rm (em)}_{\mu\nu}=-(1/4\pi)(F_{\mu\ \cdot}^{\ \
            \lambda}F_{\nu\lambda}-(1/4)g_{\mu\nu}F^{2})
\]
is the canonical energy-momentum tensor of the electromagnetic field;

\medskip
$F^{2}=F_{\alpha\beta}F^{\alpha\beta}$;

\medskip
$\phi=\ln\varphi=\ln\sqrt{-G_{55}}$ is a
new notation for the ge\-o\-met\-ri\-c scalar field;

\medskip
$\nabla^{2}=g^{\alpha\beta}\nabla_{\alpha}\nabla_{\beta}$;

\medskip
$\nabla_{\alpha}$ is the covariant derivative corresponding to $g$.

The ten equations (\ref{e4}) are the $(\mu\nu)$-projection
 of (\ref{vacuum}), the four equations (\ref{maxw}) are the
 $(\mu 5)$-projection of (\ref{vacuum}) and eq. (\ref{s4}) is the
 (55)-projection.

Thus in the present version of the KKT any vacuum solution
of the 5-dimensional Einstein equations, which is cylindrical in the
5th coordinate, corresponds to the set (\ref{e4})--(\ref{s4})
of the Einstein-Maxwell-scalar field equations in 4-dimensional GR.

\section{5-dimensional transformations and generation theorems}
            \label{gen5}

Consider the general 5-dimensional coordinate transformations
\begin{equation}\label{tr}
      x^{A'}=x^{A'}(x^{A})
\end{equation}
and choose from (\ref{tr}) those
which preserve the cylindricity condition (\ref{cyl}).
This means that the transformed metric $G_{A'B'}$
satisfies the cylindricity condition in the new coordinates $x^{A'}$:
$\partial_{5'}G_{A'B'}=0$.
To find out whether such transformations do exist, let
us formulate a theorem. The following notations will be used:
\[
\alpha^{A'}_{A} \equiv \partial x^{A'}/\partial x^{A}\]
            ---  Jacobi matrix elements;

\[ \alpha^{A}_{A'}\equiv \partial x^{A}/\partial
                        x^{A'}\]
               ---   inverse Jacobi matrix elements;
\[(\alpha^{A'}_{A}\alpha^{A}_{B'}=\delta^{A'}_{B'});\ \ \
s^{A} \equiv  \alpha^{A}_{5'}\]
---   vector-column of the inverse Jacobi matrix.
Hereafter it is assumed that a sufficiently smooth
re\-ver\-si\-ble coordinate transformation (\ref{tr}) is given.

{\bf Theorem 1} {\it The cylindricity condition (\ref{cyl}) is satisfied
in the coordinate system $x^{A'}$ if and only if the vector-column $s^{A}$
is a Killing vector of the original metric.}

{\bf Proof.}
Let the transformed metric satisfy the cylindricity condition
$G_{A'B',5'}=0$. Then the following chain of equalities takes place:
\[
G_{A'B',5'}=(G_{AB}\alpha^{A}_{A'}\alpha^{B}_{B'})_{,5'}
=G_{AB,5'}\alpha^{A}_{A'}\alpha^{B}_{B'}
+G_{AB}\alpha^{A}_{A',5'}\alpha^{B}_{B'}+
G_{AB}\alpha^{A}_{A'}\alpha^{B}_{B',5'}\]
\[
=G_{AB,C}s^{C}\alpha^{A}_{A'}\alpha^{B}_{B'}
+G_{AB}s^{A}_{,A'}\alpha^{B}_{B'}+
G_{AB}\alpha^{A}_{A'}s^{B}_{,B'}\]
\[   =(G_{AB,C}s^{C}+s^{C}_{,A}G_{CB}
+s^{C}_{,B}G_{AC})\alpha^{A}_{A'}\alpha^{B}_{B'}
=2s_{(A;B)}\alpha^{A}_{A'}\alpha^{B}_{B'}=0.
\]
Here we have used the symmetry properties of the Jacoby matrix
derivatives, the definition of $s^{A}$ and a substitution of variables.
Due to the nondegeneracy of the trans\-for\-ma\-ti\-on, the
last equality gives the Killing equations for $s^{A}$.
The necessity is proved.
If the 5-dimensional metric admits a Killing vector,
then, identifying its components with $\alpha^A_{5'}$ and repeating
the above formulae backward, we get the cylindricity condition for
the transformed metric. $\bullet$

Hereafter we will call the Killing vector that has appeared
in the theorem {\it a generating Killing vector.}
A 5-dimensional space-time can admit several
generating Killing vectors. Let us note that the conditions of the theorem
do not include the cylindricity of the original metric.

Using Theorem 1, let us prove the following useful propositions:

\begin{enumerate}
\item
{\it In the new coordinate system the vector  $\vec s$ has the
components}
$s^{A'}=\alpha^{A'}_{A}s^{A}=\alpha^{A'}_{A}\alpha^{A}_{5'}=\delta^{A'}_{5}$.
This circumstance makes clear the geometric meaning of the method
discussed in the Introduction.
\item
{\it The generating Killing covector in the new coordinate system
becomes proportional
to the vector potential of a new geometrical electromagnetic field.}
This can be easily seen from the fact that
$s^{A'}=\delta^{A'}_{5'}$. Lowering the index $A'$ using the new metric
$G_{A'B'}$, we get $s_{\mu'}=G_{5'\mu'}\sim A_{\mu'}$ (see (\ref{ident}) and
(\ref{gauge})).
This fact leads us to the hypothesis that the proposed method is related
to the Mitskievich-Horsk\'y generation procedure, which is purely
4-dimensional \cite{mitsk3}.
\item
{\it A general form of admissible transformations
is given by the expressions
\[
x^{\mu'} = F^{\mu'}(\varphi^{\nu});\ \ \
x^{5'}  =  F^{5'}(\varphi^{\nu},\varphi^{5})
\]
where $\{\varphi^{A}\}$ is a set of independent functions
defining the set of integral curves of a generating Killing vector,
$F^{A'}$ is a set of arbitrary independent
differentiable functions of its arguments.}
This proposition directly follows from the set of linear
partial differential equations obtained from the equality:
\begin{equation}\label{adc}
      s^{A'}=\alpha^{A'}_{A}s^{A}=\delta^{A'}_{5'}.
\end{equation}
The functions $\varphi^{A}$  are independent
integrals of the corresponding set of characteristics equations.
\item
The linearity of the Killing equations leads to the fact that {\it any
linear superposition of Killing vectors with constant coefficients
can be considered as a generating Killing vector.}
\item\label{rosly}
{\it The new scalar field $\varphi'$ is related to  a generating Killing
vector in the following way:
$G_{5'5'}=-\varphi^{'2}=\vec s\cdot\vec s$}.
In the new coordinate system
$\vec s\ '\cdot\vec s\ '=G_{A'B'}\delta^{A'}_{5'}\delta^{B'}_{5'}=G_{5'5'}$.
The proposition follows from the invariance of the norm.
\end{enumerate}

This last proposition allows one to get rid of the scalar field
in the transformed set of equations (\ref{e4})-(\ref{s4}),
which has been noted by Rosly. In  \cite{ros}
he has formulated and proved the following theorem:

{\bf Theorem 2 (Rosly)}
{\it If a 5-dimensional metric admits a space-like Killing vector with a constant
norm, then the admissible coordinate transformation generated by this
vector gives a 4-dimensional vacuum solution of the Einstein-Maxwell
equations.  The electromagnetic field obtained in this way satisfies the
condition  $F^{2}=0$.}

The theorem directly follows from proposition
\ref{rosly} and eqs. (\ref{e4})--(\ref{s4}) with $\phi={\rm const}$.

\section{Generation of exact solution with a scalar field}
\label{trans1}

Let us demonstrate the present procedure taking
the flat 5-dimensional metric
\begin{equation}\label{plane}
ds^2_{(5)}=dt^2-dx^2-dy^2-dz^2-d\upsilon^2.
\end{equation}
as the original one. It has a complete 15-parameter isometry group.

Let us take a generating vector in the following form:
$\vec
s=x\partial_y-y\partial_x+\alpha^{0}\partial_{t}+\alpha^3\partial_z
                        +\alpha^5\partial_\upsilon$,
which is a linear superposition of the rotation generator in the
$(xy)$ plane and translations along the $t,z,\upsilon$ axes.
Solution of eqs. (\ref{adc}) gives the following transformations
(up to 4-dimensional and gauge transformations):
\[
\begin{array}{lcr}\label{ptr}
      t &=& t'+a\upsilon';\\
      x &=& x'\sin\upsilon';\\
      y &=& y'\cos\upsilon';\\
      z &=& z'+b\upsilon';\\
      \upsilon &=& y'.
\end{array}
\]
Here $a$ and $b$ are arbitrary constants.
The metric (\ref{plane}) trans\-for\-med according to (\ref{ptr})
takes the form (we omit accents at the new coordinates):
\[
      ds^2_{(5)}=dt^2-dx^2-dy^2-dz^2-(\Delta+x^2)d\upsilon^2+2d\upsilon(a\,dt-b\,dz)
\]
where $\Delta=b^2-a^2$.
The 1-forms $\lambda$ and $A$ from (\ref{gauge}) and (\ref{ident})
have the form
\[
\lambda= \frac{1}{\sqrt{\Delta+x^2}}(a\,dt-b\,dz-(\Delta+x^2)d\upsilon);\ \
      A   = \frac{1}{2(\Delta+x^2)}(a\,dt-b\,dz).
\]
The physical metric can be found from (\ref{monada}) by the
conformal transformation (\ref{conf}). It has the form
\[
      ds^2_{(4)}=\frac{1}{X^2}[X_b\,dt^2-X(dx^2+dy^2)-X_a\,dz^2-2ab\,dt\,dz],
\]
where $X=\Delta+x^2$; $X_a=x^2-a^2$; $X_b=x^2+b^2$.
Taking an external differential of the 1-form $A$, we get the
electromagnetic field 2-form $F$:
\[
      F=dA=\frac{x}{(\Delta+x^2)^2}(adt\wedge dx+bdx\wedge dz).
\]
The electric current density 4-vector (\ref{maxw}) is
\[
      \vec j=-\frac{3}{4\pi}\phi_{,\alpha}F^{\alpha\mu}\partial_{\mu}=
            -\frac{3}{4\pi}(a\partial_t+b\partial_z).
\]
Assuming that the electric charge is carried
by the ge\-o\-me\-tri\-zed matter,
i.e. $\vec j=\sigma\vec u_{c}$ where $\sigma$ is the electric
charge density and $\vec u_{c}$ is the 4-velocity of charged matter, we get:
\[
      \vec j\cdot\vec j=\sigma^2=-\frac{9\Delta}{16\pi^2X^2};\ \ \
            \sigma=\frac{3}{4\pi X}\sqrt{-\Delta};\quad
      \vec u_{c}=\vec j/\sigma=-\frac{X}{\sqrt{-\Delta}}(a\partial_t+b\partial_z).
\]
From the expression for $\sigma$ one can get the following
restriction on the parametres $a$ and $b$: $\Delta<0$, or $|a|>|b|$.

\end{document}